\documentclass[runningheads]{llncs}
\usepackage{graphicx}
\usepackage{algorithmic}
\usepackage{algorithm}
\usepackage{amsmath, amssymb, amsfonts}
\usepackage{array}
\usepackage{xcolor}
\usepackage[caption=false,font=normalsize,labelfont=sf,textfont=sf]{subfig}
\usepackage{textcomp}
\usepackage{stfloats}
\usepackage{url}
\usepackage{verbatim}
\usepackage{cite}
\usepackage{mathtools}
\usepackage{multirow, multicol}
\usepackage{tabularx}
\usepackage{ragged2e}
\usepackage{blindtext}
\usepackage{colortbl}
\usepackage{hyperref}
\usepackage{fontawesome5}  

\newcommand{\orcid}[1]{\href{https://orcid.org/#1}{\textcolor[HTML]{A6CE39}{\faOrcid}}}

\begin{document}
\title{Architecture for Protecting Data Privacy in Decentralized Social Networks\thanks{supported by the Erasmus+ Programme International Credit Mobility for studies supported by the University of the Aegean through EU funds. Additionally, it received support from the Australian Research Council through the Discovery Project under Grant DP200100731. The collaboration involved the University of the Aegean (Greece) and the Royal Melbourne Institute of Technology (Australia).}}
%
%
\author{Quang Cao\inst{1}(\faEnvelope) \orcid{0000-0001-9649-943X} \and
Katerina Vgena\inst{2}\orcid{0000-0001-7619-4014} \and
Aikaterini-Georgia Mavroeidi\inst{2}\orcid{0000-0002-3270-880X} \and
\\ Christos Kalloniatis\inst{2}\orcid{0000-0002-8844-2596} \and
Xun Yi\inst{1}\orcid{0000-0001-7351-5724} \and
Son Hoang Dau\inst{1}\orcid{0000-0002-2276-017X}}
\authorrunning{Q. Cao et al.}
%
\institute{RMIT University, Melbourne, Victoria, Australia \\
\email{\{nhat.quang.cao2, sonhoang.dau, xun.yi\}@rmit.edu.au}
\and
Privacy Engineering and Social Informatics Laboratory, Department of Cultural Technology and Communication, University of the Aegean, Mytilene, Greece\\
\email {\{kvgena, kmav, chkallon\}@aegean.gr}}

\maketitle       

\begin{abstract} 
Centralized social networks have experienced a transformative impact on our digital era communication, connection, and information-sharing information. However, it has also raised significant concerns regarding users’ privacy and individual rights. In response to these concerns, this paper proposes a novel Decentralized Social Network employing Blockchain technology and Decentralized Storage Networks completed by Access Control Smart Contracts. The initial phase comprises a comprehensive literature review, delving into decentralized social networks, explaining the review methodology, and presenting the resulting findings. Building upon these findings and an analysis of previous research gaps, we propose a novel architecture for decentralized social networks. In conclusion, the principal results highlight the benefit of our decentralized social network to protect user privacy. Moreover, the users have all rights to their posted information following the General Data Protection Regulation (GDPR).

\keywords{Decentralized Social Networks \and Blockchain \and Decentralized
Storage Network \and Shamir Secret Sharing \and Smart Contract \and Privacy.}
\end{abstract}
\section{Introduction}
\label{sec:intro}

Privacy in systems is crucial for safeguarding personal information and fostering trust in digital interactions in each system \cite{mavroeidi2019interrelation} and several approaches have been published to address such issues \cite{kalloniatis2008addressing}.
Privacy concerns have become paramount in the contemporary digital landscape, especially in social networks \cite{garton1997studying} where centralized entities often wield significant control over user data \cite{houghton2014privacy,vgena2022determining}. This research critically examines the existing challenges surrounding privacy in social networks. It proposes an innovative solution by integrating decentralized architectures with a specific focus on blockchain technology \cite{nofer2017blockchain}. The study explores how decentralization, facilitated by blockchain, can redefine the landscape of social networking by redistributing control and ownership of user data. Blockchain, renowned for its secure and transparent nature, is leveraged as a foundational technology to underpin a novel social network architecture \cite{zheng2018blockchain}. This research addresses the pressing issue of privacy in social networks and contributes by providing an architecture for protecting privacy in social networks. Social networks have been pivotal in shaping digital communication, connection, and information-sharing in our contemporary era \cite{borgatti2018analyzing, butler2020social, oseni2018instant}. However, their impact has been accompanied by growing concerns about users' privacy \cite{zhang2010privacy}. 
In response to these apprehensions, our study defines attributes that characterize Decentralized Online Social Networks (DOSNs) \cite{datta2010, de2017, choi2020, rahman2019}. The advantages of DOSNs highlight their absence of a central authority, user-controlled information, and the potential of blockchain technology in this context. DOSNs offer an alternative to centralized Online Social Networks (OSNs), emphasizing distributed implementation through models like peer-to-peer architectures \cite{datta2010, de2017}. 

The research question, \textit{What are the Decentralized social networks?} guides the inquiry. The study utilizes a Boolean approach to specify keywords and examines how these variables contribute to defining DOSNs. The search strategy involves combining terms such as (Decentralized OR Distributed) AND (Social Media OR Social Networks) AND (Blockchain) AND (Definition). The study covers recent publications, considering the innovative nature of the topic. The review focused on titles and abstracts of journal articles, book chapters, workshop papers, and conference papers to ensure relevance to the study's objectives. The study utilized a Prisma 2009 flow diagram \cite{PRISMA} to represent the reviewing process visually, adapting the PRISMA methodology to align with the study's rationale. The review identified a gap in defining DOSNs and emphasized the need for a legal framework, particularly within the EU's General Data Protection Regulation (GDPR). The text suggests that personal data is a crucial aspect of online representation, and GDPR distinguishes \textit{sensitive data}, outlining specific processing conditions for its protection. DOSNs leverage decentralized networks, giving users greater control over their information. The text underscores the importance of addressing legal considerations, particularly within the GDPR framework, to protect privacy and security in the evolving landscape of decentralized online social media.

Moreover, this paper proposes an innovative approach to social networking by integrating Blockchain technology and Decentralized Storage Networks, complemented by Access Control Smart Contracts. The proposed architecture ensures that users retain control and ownership rights over their posted information, aligning with the General Data Protection Regulation (GDPR) principles \cite{voigt2017eu}. Following the distinct types of personal data in GDPR, the researchers distinguished sensitive data as it is to be subjected to specific processing conditions. The initial phase of this research involves an in-depth review of decentralized social networks \cite{datta2010}, elucidating the methodology employed for the literature review and presenting the resulting findings.

In the proposed solution for a Decentralized Online Social Network (DOSN), user data is decentralized and stored in a distributed ledger to eliminate a centralized point of failure. Blockchain technology creates an immutable data record, fostering transparency and preventing unauthorized modifications without unanimous user consent. This ensures data authenticity and security while addressing data gaps. However, user data is stored in plaintext, potentially compromising anonymity, and some sensitive information has restricted access. A blockchain-based access control system is introduced, following a role-based access control (RBAC) model to prevent unauthorized users. Access control policies are stored on the blockchain, providing public auditability and allowing verification of user rights even when the data owner is not actively using the network. Users execute transactions through smart contracts to manage access rights, assigning roles to authorized individuals. However, the framework relies on a designated group of trusted nodes to store data, ensure availability without the resource owner, and verify access control for user requests.
Hence, our study proposes a novel architecture for decentralized social networks to address the identified challenges, safeguarding users' privacy. Through integrating blockchain and decentralized storage networks \cite{karaarslan2020data}, our proposed solution responds to the pressing privacy concerns in social networks. It establishes a secure and user-centric foundation for the future of digital communication and information sharing.

Our main contributions are summarized below.
\begin{itemize}
\item  We reviewed the literature comprehensively on Decentralized Online Social Networks (DOSNs) using the PRISMA methodology \cite{page2021prisma}. Through this review, we delineated the architectures of these DOSNs and highlighted their recent limitations.
\item We introduced a novel architecture for DOSNs designed to address the shortcomings identified in the existing architectures.
\end{itemize}

The structure of the paper is outlined as follows. In Section \ref{sec:Rprocess}, an overview of the review process is provided, detailing the methodology employed by the researchers and presenting the outcomes. The authors introduce their inventive architecture for decentralized online social networks in Section \ref{sec:Architecture}. The contribution of the research is discussed in Section \ref{sec:discusstion}, and the paper concludes in the final Section \ref{sec:conclusion}, encompassing the authors' concluding remarks, acknowledging the limitations of the presented work, and suggesting potential directions for future developments.

\section{Review process}
\label{sec:Rprocess}

\subsection{Method}
\label{sec:Method}

One of the primary concerns of this study is to specify proper attributes that constitute the characterization, while providing a definition of the decentralized online social networks (DOSNs). Therefore, the research question \textit{What are the Decentralized social networks?} was formulated.

Under this spectrum, our study specifies the necessary keywords using the Boolean approach. In other words, we can examine how the variables above are applied in defining the Decentralized social networks. The documents were gathered from various sources using a search string created by combining the following terms and connecting them with the Boolean OR/AND operators. More precisely, the Table \ref{tab:SearchStrategy} summarizes the keywords applied per database:

\textbullet \hspace{1mm} Search terms for RQ: (Decentralized OR Distributed) AND (Social Media OR Social Networks) AND (Blockchain) AND (Definition)\

The study was conducted during spring 2023, covering recent publications as the topic of interest is quite innovative. The time period was set up untill  June 1st, 2023. No starting date was set as researchers would include papers that include the search terms regardless of their publication time. More precisely, the review was conducted on academic databases in, IEEEXplore, Scopus, ScienceDirect, Web of Science and ACM Digital Library. The search was conducted on the titles and abstracts of journal articles, book chapters, workshop papers, and conference papers to verify their relevance to the objectives of this study.
In this systematic review, we followed a specific search strategy to describe the steps that the study followed:

\vspace{-10pt}
\begin{table}[ht]
\caption{Search Strategy}
  \begin{tabularx}{\textwidth}{X|X} 
    \hline
    \multirow{5}{*}{\textbf{Academic databases searched}} & IEEExplore\\ 
    & Scopus\\ 
    & Science Direct\\ 
    & ACM Digital Library\\ 
    & Web of Science \\ 
    \hline
    \multirow{4}{*}{\textbf{Target Items}} & Journal Papers\\ 
    & Workshop Papers\\ 
    & Conference Papers\\ 
    & Book Chapters \\ 
    \hline
    \multirow{2}{*}{\textbf{Search applied to}} & Titles\\ 
    & Abstracts\\ 
    \hline
    \textbf{Language} & English\\ 
    \hline
    \multirow{2}{*}{\textbf{Publication Period}} & Start date not set\\ 
    & End date until June 1, 2023\\ 
    \hline
  \end{tabularx}
  \vspace{5pt}
  
  \label{tab:SearchStrategy}
\end{table}
\vspace{-20pt}

Given the extensive volume of results generated by a broad search, and to maintain a manageable scope, the number of outcomes was constrained by applying specific inclusion and exclusion criteria outlined in Table \ref{tab:IncExcCriteria}. First, to ensure the effectiveness of comprehending this research, it was necessary for the studies to be authored in English. Next, full-text articles were selected and duplicates were excluded. 

\begin{table}[ht]
\caption{Inclusion and Exclusion Criteria}
  \begin{tabularx}{\textwidth}{X|X} 
    \hline
    \textbf{Inclusion Criteria} & \textbf{Exclusion Criteria} \\ 
    \hline
    Written in English & Written in other languages \\ 
    \hline
    Full-text articles & Non-full text access to the articles \\ 
    \hline
    Check for duplicates & Exclude duplicates \\ 
    \hline
  \end{tabularx}
  \vspace{-10pt}
  \label{tab:IncExcCriteria}
\end{table}

More specifically, as indicated by the results presented in Table \ref{tab:DbReview}, the majority of them were discovered within the ACM Digital Library and Scopus databases, with only a small portion of the results originating from the ScienceDirect and Web of Science databases.

\begin{table}[ht]
\caption{Databases for the Review}
  \begin{tabularx}{\textwidth}{X|X} 
    \hline
    \textbf{Databases} & \textbf{Findings per Keyword} \\ 
    \hline
    IEEE Xplore & Abstract 7, Title 7\\
    \hline
    ScienceDirect & Title and Abstract 4\\
    \hline
    Web of Science & Title and Abstract 4\\
    \hline
    ACM Digital Library & Abstract 37, Title 1\\
    \hline
    Scopus & Title and Abstract 16\\
    \hline
  \end{tabularx}
  \vspace{-10pt}
  \label{tab:DbReview}
\end{table}

\begin{figure}[htb!]
\vspace{-55pt}
    \centering
    \includegraphics[scale=0.29]{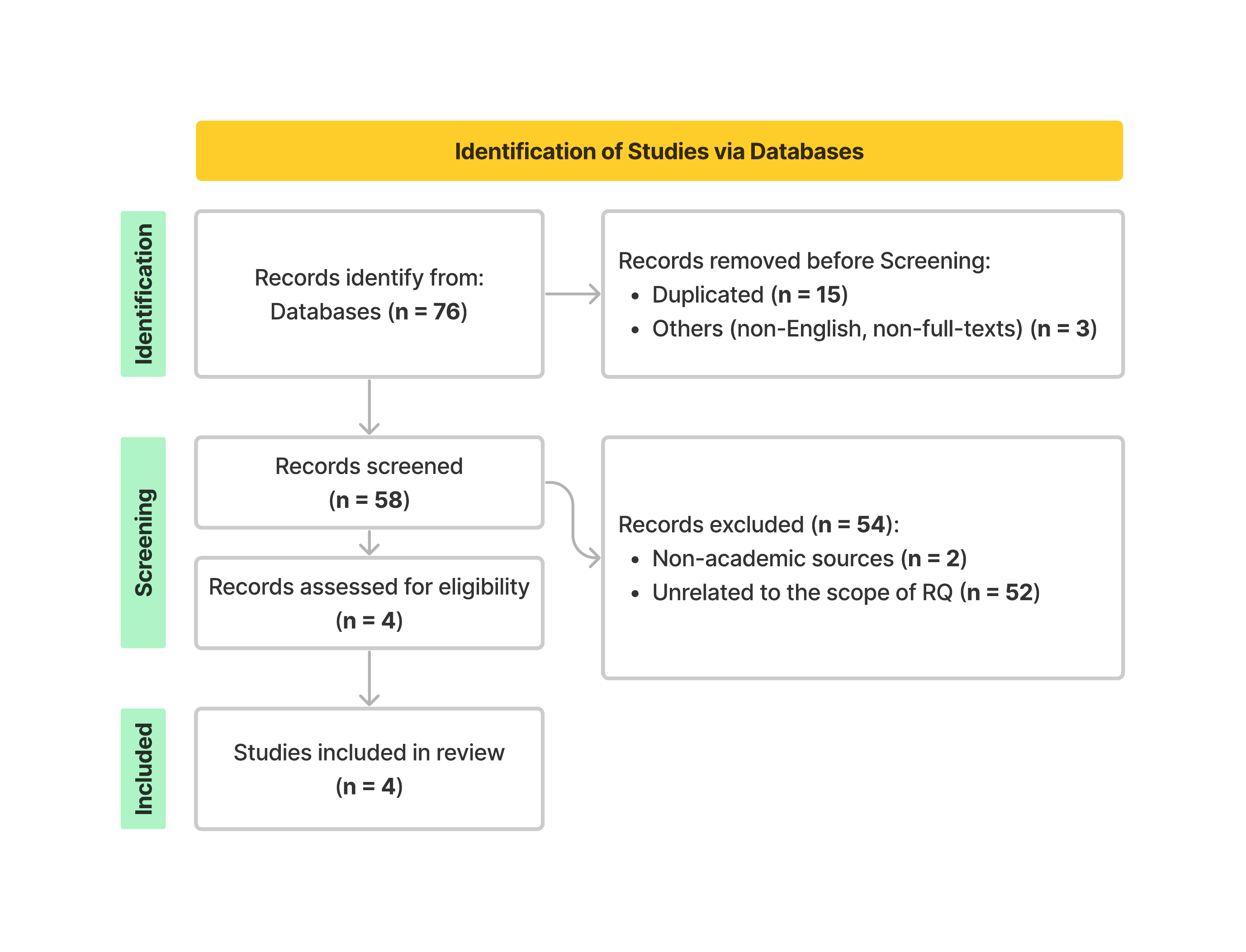}
    \vspace{-45pt}
    \caption{Prisma Screening Process, Systematic Literature Review - flow diagram}\
    \label{fig:Prisma}
\end{figure}
\vspace{-15pt}

A Prisma 2009 flow diagram was also used to represent the reviewing process visually \cite{page2021prisma}. Figure \ref{fig:Prisma} illustrates the screening process proposed by PRISMA methodology with the necessary adjustments of the proposed visualization to match the study's rationale. It is worth highlighting that numerous studies were encountered; however, a significant proportion of them did not align with the criteria outlined in the relevant Table and were, therefore, unsuitable for this research. A total of 76 studies were initially identified, and following the removal of duplicates, non-English and non-full-text studies, 58 studies were subject to screening. Out of these, 54 studies were excluded as they were unrelated to the scope of RQ, and one was not an academic study. Considering the above steps, four papers were selected and included in the study. 

In the table below, an analysis of the findings based on the applied databases is presented. For example, in the IEEExplore database, 14 studies were found, all of which were in English and constituted full-text articles, in contrast to the ACM database, where one of the articles was not full-text. Subsequently, duplicate entries were removed, and the final step involved eliminating those unrelated to the blockchain and social media or including a definition for decentralised social networks.

\vspace{-10pt}
\begin{table}
  \centering
  \caption{Selected Papers}
  \begin{tabular}{m{0.27\linewidth}|m{0.15\linewidth}|m{0.1\linewidth}|m{0.1\linewidth}|m{0.1\linewidth}|m{0.12\linewidth}|m{0.1\linewidth}} 
    \hline
    \textbf{Exclusion criteria} & \textbf{Total Number of articles} & \textbf{IEEE Xplore} & \textbf{Science Direct} & \textbf{Web of Science} & \textbf{ACM Digital Library} & \textbf{Scopus} \\ 
    \hline
    \textbf{Total Number with the keywords} & 76 & 14 & 4 & 4 & 38 & 16\\ 
    \hline
    \textbf{Written in English} & 75 & 14 & 4 & 4 & 38 & 15\\ 
    \hline
    \textbf{Full text-articles} & 74 & 14 & 4 & 4 & 37 & 15\\ 
    \hline
    \textbf{Exclude duplicates} & \multicolumn{6}{c}{\textbf{58}}\\
    \hline
    \textbf{Exclude Papers that do not discuss blockchain and social media and not Defining DOSN} & \multicolumn{6}{c}{\textbf{54}}\\
    \hline
  \end{tabular}
  \vspace{-10pt}
  \label{tab:Selected Papers}
\end{table}

After proceeding with that necessary step, this study has also applied a second phase of exclusion criteria for choosing the reviewed papers. More precisely, the researchers, after the screening process also excluded the papers that did not discuss blockchain and social media applications under the DOSNs spectrum in a way that matched their scope of research; thus, they were able to finalize the number of articles for RQ. The studies that were chosen for conducting this review are presented in the Table \ref{tab:selectedPapers}.

\begin{table}[ht]
  \centering
  \caption{Selected Papers.}
  \begin{tabularx}{\textwidth}{X|X}
    \hline
    \textbf{Authors} & \textbf{Title} \\
    \hline
    Ur Rahman, Mohsin; Guidi, Barbara; Baiardi, Fabrizio & Blockchain-based access control management for Decentralized Online Social Networks \\
    \hline
    Guidi, Barbara; Conti, Marco; Passarella, Andrea; Ricci, Laura & Managing social contents in Decentralized Online Social Networks: A survey \\
    \hline
    Rahman, M.U.; Baiardi, F.; Guidi, B.; Ricci, L. & Protecting personal data using smart contracts \\
    \hline
    Guidi, Barbara; Clemente, Vanessa; García, Tomás; Ricci, Laura & A Rewarding Model for the next generation Social Media \\
    \hline
  \end{tabularx}
  \vspace{-15pt}
  \label{tab:selectedPapers}
\end{table}

\subsection{Results}
\label{sec:Results}

Decentralized Online Social Networks do not have a service provider that acts as a central authority, and users have more control over their information \cite{URRAHMAN202041}. Blockchain technology has emerged as a potential solution for addressing this issue \cite{guidi2020blockchain}. However, in Decentralized Online Social Networks, blockchain implementations typically function as a storage mechanism, resulting in publicly accessible content, including users' social data \cite{rahman2020blockchain}. This enables individuals to manage their data independently and reduces reliance on large service providers for controlling personal information. Additionally, DOSNs facilitate customization options and enable content creators to be rewarded directly without the involvement of central authorities. By leveraging the decentralized nature of DOSNs, users can exercise greater control over their personal information and participate in a more equitable content ecosystem \cite{URRAHMAN202041}.\\

More specifically, a Decentralized Online Social Network (DOSN), where there is no single controlling authority or service provider, is an alternative to centralized Online Social Networks (OSNs) like Facebook or Twitter, where there is no single controlling authority or service provider \cite{rahman2019}. In a DOSN, the network is implemented in a distributed manner \cite{rahman2019}, utilizing various models such as peer-to-peer (P2P) architectures or networks of trusted servers \cite{guidi2018managing}. As mentioned previously, the decentralization of data storage allows users to have greater control over their personal information.  Furthermore, the fourth paper defines a reward system that incentivises and engages existing and potential users using the HELIOS Platform \cite{10.1145/3411170.3411247}. HELIOS is a decentralized social media platform that addresses the dynamic nature of human communications, and it includes techniques such as decentralisation, context detection in IoT environment, real and virtual object networking, peer-to-peer based content streaming and validation \cite{10.1145/3411170.3411247}.

Therefore, after conducting the review process, we can conclude with a first attempt at defining the DOSNs. More precisely, based on the selected papers \cite{URRAHMAN202041,guidi2018managing,rahman2019,10.1145/3411170.3411247}, Decentralized Online Social Networks (DOSNs) are Online Social Networks implemented by exploiting decentralized networks. They are an alternative to centralized Online Social Networks (OSNs) in the sense that the network is implemented in a distributed manner with no single central controlling authority or service provider, and users have more control over their personal information.

During the review, the researchers were able to identify the research gap in defining decentralized online social media while at the same time providing the necessary Legal Framework for protecting privacy and security concepts in the DOSNs. Therefore, it became evident that there should be a short analysis of the EU legal framework as personal data plays a vital role in representing oneself online, engaging audiences and establishing meaningful interactions on Social Media (SM). The term \textit{personal data} has been described thoroughly after the implementation of the General Data Protection Regulation (GDPR) \cite{RUGHINIS2021105585}.
Social media can prove to be the \textit{melting pot} of personal data. What we mean here is that according to the GDPR, different categories of personal data are distinctively presented in order to be subjected to specific processing conditions. More precisely, GDPR clearly distinguishes \textit{sensitive data} as a category that includes the revelation of users’ characteristics, such as racial or ethnic origin, political opinions, religious or philosophical beliefs, trade union memberships, genetic data, biometric data processed solely to identify a human being, health-related data, and data concerning a person’s sex life or sexual orientation \cite{europaWhatPersonal}. Namely, Article 4(13), (14) and (15)  and Article 9 and Recitals (51) to (56) of the GDPR are descriptive of the category of \textit{sensitive data} in a sense that they require special processing conditions\cite{europaWhatPersonal,Your}.

\section{Architecture for Decentralized social networks}
\label{sec:Architecture}

\begin{figure}[htb!]
\centering
\includegraphics[scale=0.72]{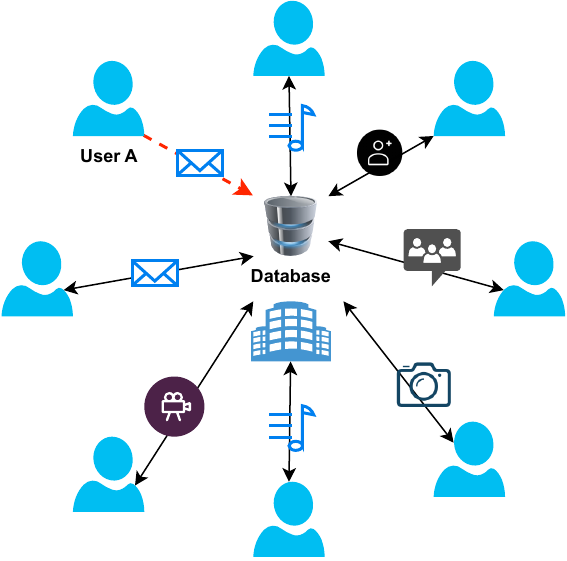}
\vspace{-10pt}
\caption{The traditional OSN architecture \cite{choi2020}.}
\label{fig:osn}
\end{figure}

\begin{figure}[htb!]
\centering
\includegraphics[scale=0.72]{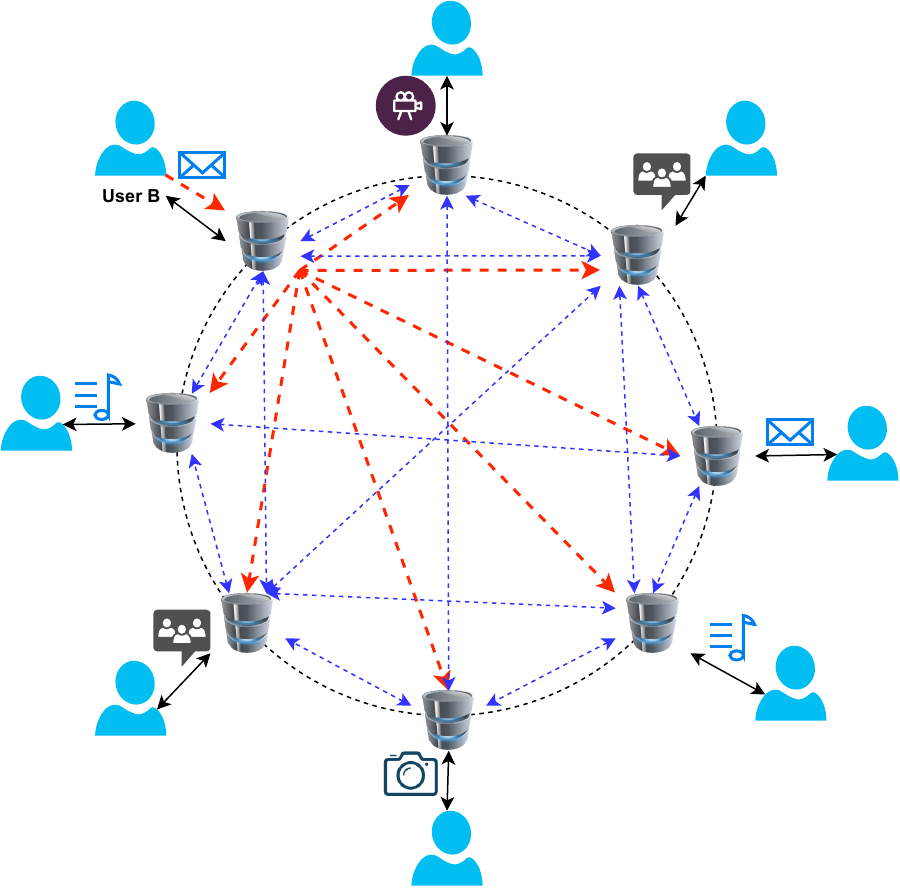}
\caption{The DOSN architecture \cite{choi2020}.}
\label{fig:dosn}
\end{figure}

A Decentralized Online Social Network (DOSN) is an OSN that operates on a distributed platform, fostering collaboration among multiple independent users utilizing the OSN, such as a peer-to-peer (P2P) network \cite{datta2010, de2017}. Unlike centralized OSNs, where the service provider manages user data (see Fig. \ref{fig:osn}), DOSNs employ a decentralized structure in decentralized managers (see Fig. \ref{fig:dosn}). 

\vspace{-15pt}

In Figure \ref{fig:osn}, users interact with the platform by posting messages or retrieving information. The central database stores and manages all user data. For instance, if User A shares a new message, the platform can censor it. Other users can exclusively access the new message through the platform. In the event of a central database attack, the security of all users' data is compromised.

\vspace{-15pt}

To solve centralized point of failure, in Figure \ref{fig:dosn}, user data is decentralized and stored within a distributed ledger. When User B shares a new message, the entire blockchain system is notified, as the red dotted lines indicate. Utilizing blockchain technology establishes an immutable data record that can be publicly accessible, promoting data transparency, as evident in the blue dotted lines. Modification of information within this architecture is impossible without unanimous consent from all users, ensuring freedom of speech. This setup addresses data gaps, providing both data authenticity and security. However, all parties have access to user data stored in plaintext. While users can maintain anonymity, their information may be gathered and linked to real identities. Additionally, some data is sensitive, with access restricted to authorized users.

\begin{figure}[htb!]
\centering
\includegraphics[scale=0.8]{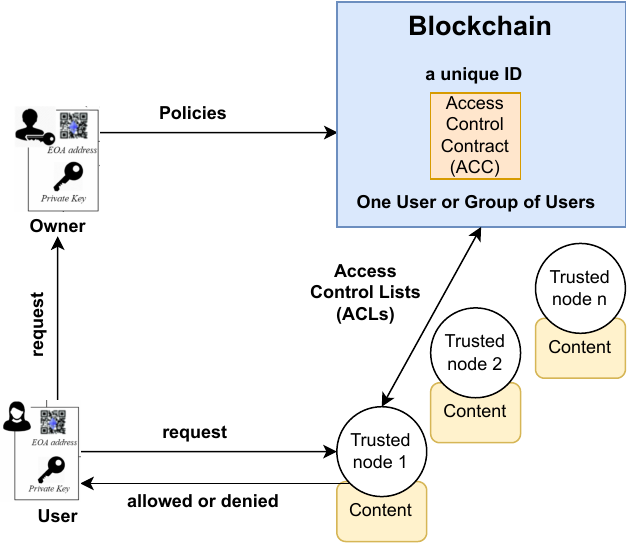}
\caption{The DOSN architecture with Access Control Contract (ACC) stored in Blockchain and Users' contents stored on Trusted nodes.}
\label{fig:DOSNPre}
\end{figure}

The prevention of unauthorized access to an individual owner's data in a Decentralized Online Social Network (DOSN) is achieved through access control. Rahman et al., \cite{rahman2019} introduced a blockchain-based access control system for DOSNs. The access control policies, based on role-based access control (RBAC) (see Fig. \ref{fig:DOSNPre}), are stored on the blockchain, ensuring public auditability and enabling verification of user rights even when the data owner is not actively logged into the social network. Its decentralized access control management relies on DOSN users to execute transactions through smart contracts, allowing them to grant, revoke, or update access rights. A unique address (a unique public key) uniquely identifies each user in the DOSN. Emphasizing the Role-based Access Control Model (RBAC), which is crucial in DOSN content management, users can utilize transactions to assign roles to colleagues, family, friends, etc., ensuring that only authorized users have access to the resources of the data owner. However, this architecture requires a designated group of trusted nodes responsible for storing data, ensuring data availability even in the absence of the resource owner. These trusted nodes are crucial in verifying access control for any user request. Drawing upon the findings presented in Table \ref{tab:selectedPapers}, we encapsulate the primary contributions of these works in Figure \ref{fig:DOSNPre}. Critical problems identified involve trusted nodes, the associated storage expenses, and the users' data being stored in plaintext, particularly in scenarios where each trusted node must replicate the content. If a trusted node turns malicious, the compromise of a trusted node poses a significant threat, potentially causing the entire system to collapse. In such instances, the malicious node gains access to and can manipulate user data. Furthermore, the compromised node enables unrestricted access to read and modify stored content. The malicious node's ability to deceive users is heightened since data integrity cannot be verified. Additionally, considering each trusted node storing identical content in scenarios with $k$ trusted nodes and $n$ contents, the cumulative storage cost rises to $k \times n$. 

\subsection{Our architecture}
\label{sec:OurArc}

\begin{figure*}[htb!]
\centering
\includegraphics[scale=0.5]{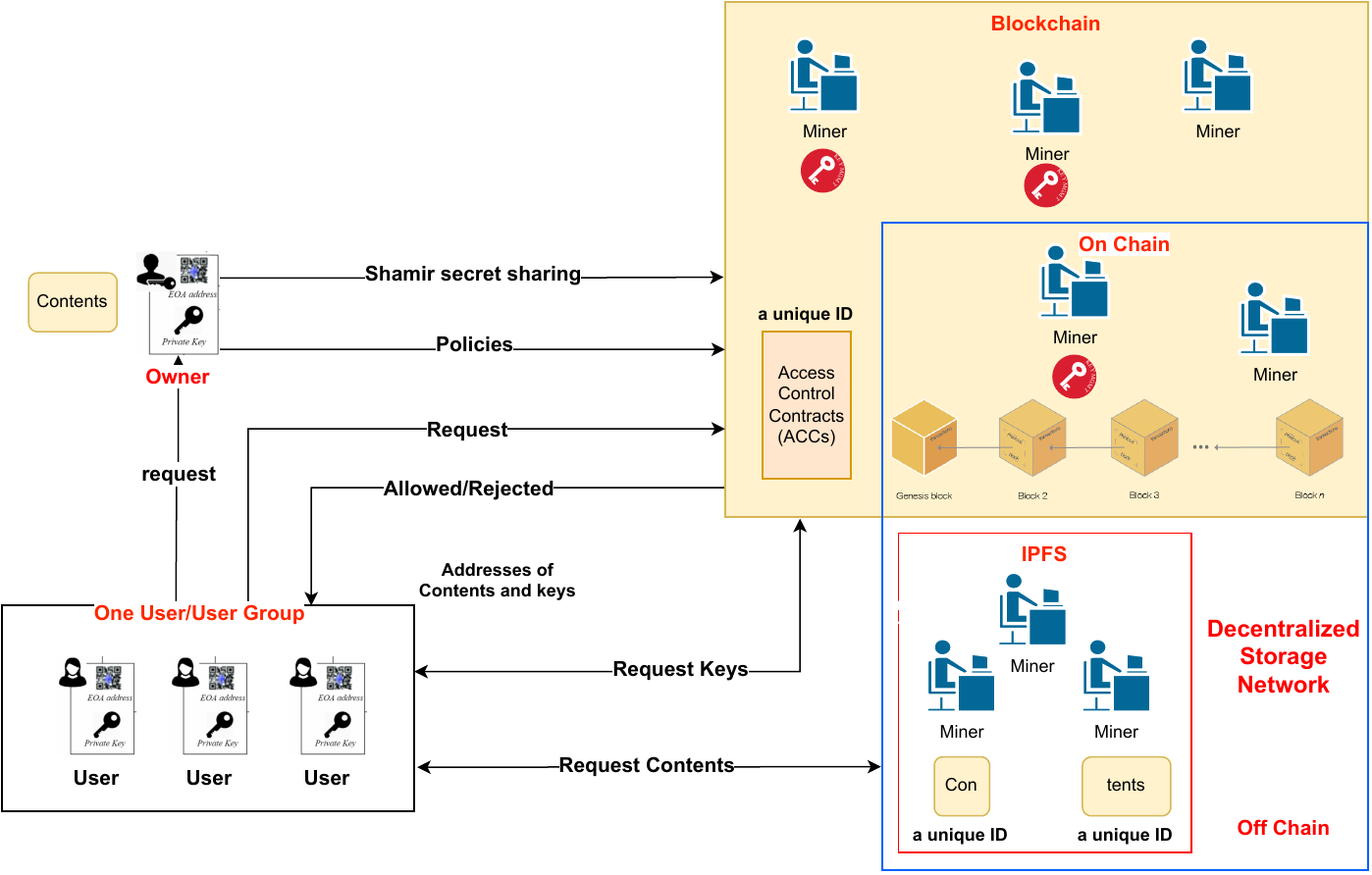}
\caption{DOSN Architecture.}
\label{fig:ourDOSN}
\end{figure*}

In Figure \ref{fig:ourDOSN}, we introduce a novel Distributed Online Social Network (DOSN) architecture addressing previous drawbacks. At a high level, content owners employ encryption to safeguard their sensitive data, distributing individual parts of each piece to a decentralized storage system like Filecoin \cite{Filecoin2017, benet2014}. Decryption keys undergo fragmentation through Shamir Secret Sharing \cite{sohrabi2020} to enhance security, with segments stored across various miners. Owners establish policies and smart contracts for automated verification of access control lists based on user requests. Upon successful validation against the Access Control Contracts, users receive a list of miners holding the corresponding key parts and unique IDs for desired content downloads. By obtaining enough key parts and content information, users reconstruct the decryption key and decrypt the received content. Merkle Directed Acyclic Graphs \cite{merkleDAG} ensures content integrity, enabling users to verify decrypted content even if a storage node turns malicious. In this scenario, lacking comprehensive security keys, the malicious node cannot alter or deceive users. Any attempted modifications to content parts can be verified by comparing the stored root of the Merkle Directed Acyclic Trees, stored in the Blockchain, providing users with an immutable record of the content structure.

\textbf{Decentralized Storage Network: }A decentralized storage network, exemplified by Filecoin \cite{Filecoin2017}, constitutes a storage infrastructure where data is distributed across numerous nodes in a network, eliminating the necessity for a central authority or a singular point of control. In Filecoin \cite{Filecoin2017}, network robustness is attained through the replication and distribution of content, with automatic detection and repair of replica failures. Clients can choose replication parameters to safeguard against various threat models. Operating as an incentive layer on top of IPFS \cite{benet2014}, Filecoin offers storage infrastructure for diverse data types. Its advantages lie in decentralizing data, facilitating the development and operation of distributed applications, and supporting the implementation of smart contracts.

\begin{figure*}[htb!]
\centering
\includegraphics[scale=0.55]{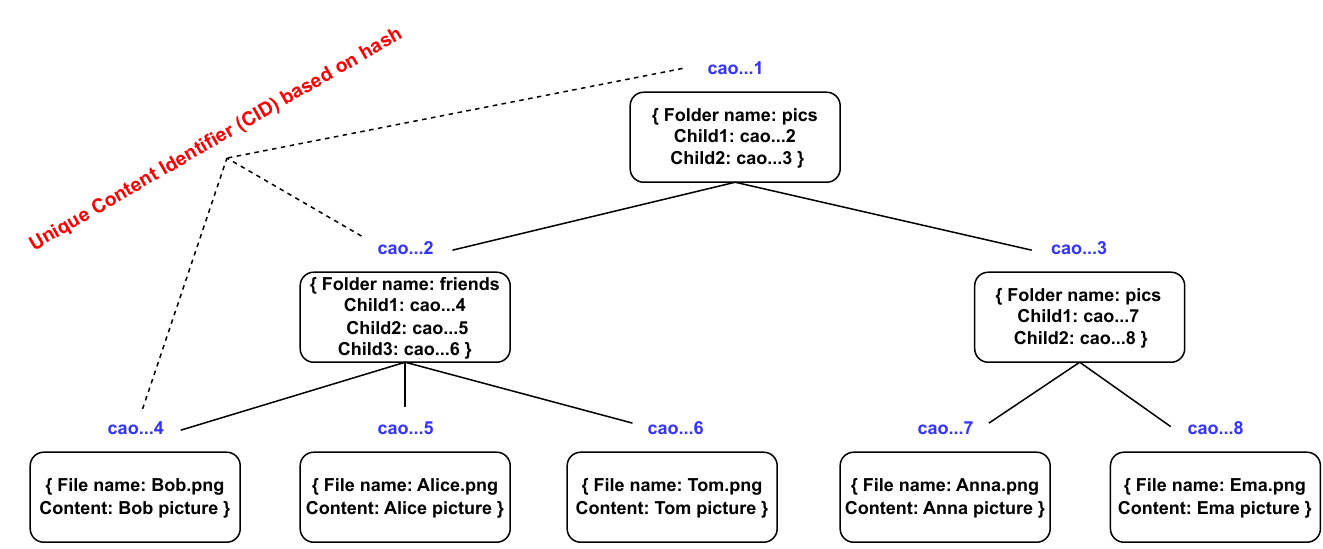}

\caption{Merkle Directed Acyclic Graphs (DAGs) \cite{merkleDAG}.}
\label{fig:DAGs}
\end{figure*}

IPFS \cite{benet2014} used Merkle Directed Acyclic Graphs (DAGs) (see Fig. \ref{fig:DAGs}) to represent a graph structure characterized by directionality and acyclic properties. In a directed graph, edges have a specific direction, such as the relationship between a directory and its contained file, where the file does not encompass its directory. The acyclic nature ensures the absence of loops within the graph. Notably, the length of a Content Identifier (CID) in this context is determined by the cryptographic hash of the underlying content rather than the content's size, adding a layer of security and efficiency to the representation of data in the graph.

Distributing files in the DAGs is open to anyone interested in contributing. Nodes from diverse geographical locations can actively participate in serving the data, fostering a globally distributed network. Each Directed Acyclic Graph (DAG) segment possesses its unique Content Identifier (CID), enabling independent distribution. This design facilitates the straightforward identification of alternative providers for the same data, promoting redundancy and resilience. The nodes forming the DAG are compact, allowing for simultaneous downloads from various providers enhancing efficiency. Moreover, larger datasets can seamlessly incorporate the original dataset by linking it as a child within a broader DAG structure, streamlining the management of expansive data sets. Deduplication is an additional characteristic of Merkle Directed Acyclic Graphs, allowing for the recycling of data and meticulous tracking of all versions of modified data.

In Fig. \ref{fig:ourDOSN}, the contents of a user are distributed across various segments in multiple miners. Each miner exclusively retains a portion of the encrypted user's contents off-chain. On the Blockchain, only the root of a Merkle Directed Acyclic Graph (DAG) is stored on-chain. The user can use the root and proof from a storage miner to verify their retrieved data later.

\textbf{Shamir Secret Sharing Scheme: }Due to the encryption of data by its owner, we have proposed an efficient approach to share decrypted keys on the Blockchain. Our method employs the Shamir Secret Sharing Scheme for managing encryption keys, similar to the approach presented in \cite{sohrabi2020}. However, in contrast to storing encrypted data on the Cloud, our approach involves storing encrypted data on a Distributed Storage System.

The Shamir Secret Sharing Scheme, a cryptographic technique introduced by Adi Shamir in 1979 \cite{shamir1979}, functions by distributing a secret into multiple shares or parts, requiring a minimum threshold of shares for reconstructing the original secret. This scheme utilizes polynomial interpolation to generate a set of points on a polynomial, each representing a share. In Fig. \ref{fig:ourDOSN}, we see that some Miner nodes will store a decryption key share.

When a user wants to fetch specific content from the owner (his friend), he must pass an Access Control Contract (ACC) that the owner created and stored on the Blockchain.

\textbf{Access Control Contracts (ACCs)} refer to a mechanism employed in blockchain and decentralized systems to regulate and manage access to specific resources or information. These contracts are created and stored on the blockchain and define the conditions and rules by data owners. Access Control Contracts ensure secure and authorized interactions within decentralized networks. The Access Control Contracts (ACCs) in our system encompass five primary functions:

Policy Creation: Owners can create policies for specific content, defining access rights. To do so, Owners must provide their address, access permissions to Access Control Lists (ACLs), addresses of Miners holding key parts, and Content Identifiers (CIDs) of the content.

Policy Update: Owners can update any policy at any time by initiating a policy update transaction directed to the ACC smart contract.

Policy Revocation: Owners can promptly revoke a policy from the policy list by sending a policy revocation transaction to the ACC smart contract.

Access Control: The ACC contract facilitates verifying access permissions on the blockchain when a user requests access. This function enables the ACC to verify access permissions on the blockchain. If the user is in the ACLs, they will receive addresses of Miners holding key parts and CIDs of the content. Subsequently, the Client requests the Decentralized Storage Network to download their desired content and its Merkle proof. The Client may also request Miners to download key parts. After confirming the correctness of the retrieved content, the Client can reconstruct the private key and decrypt the encrypted content.

ACC Deletion: Owners can deactivate the ACC smart contract by initiating a contract deactivation transaction, freeing up the storage occupied by the ACC code.

\section{Discussion}
\label{sec:discusstion}
This research highlights a critical problem in the realm of social networks, specifically addressing the challenges posed by centralized and decentralized platforms in terms of privacy and security \cite{qamar2016centralized,de2018survey}. Centralized networks, exemplified by Facebook, have been a source of concern due to the central entity's control over user data, potentially leading to exploitation without explicit consent \cite{zhang2010privacy}. On the other hand, decentralized networks, while aiming to empower users with greater control, introduce complexities such as data security management and reliance on blockchain technology \cite{bahri2018decentralized}. The identified research gap revolves around the limitations of existing decentralized online social networks, particularly in the implementation of blockchain as a storage mechanism. Two key issues are underscored: the dependence on Trusted nodes, creating vulnerability to system compromise, and the challenge of users trusting the accuracy of content without tampering. 

To bridge this gap, the proposed architecture introduces a novel approach utilizing a Decentralized Storage Network. This research contribution stands as a significant step towards enhancing the robustness and user-centric security of decentralized social networks. One of our findings concerns the identification of privacy problems in both centralized and decentralized social networks to record the possible differences \cite{de2018survey,taheri2015security}. Centralized social networks pose significant privacy challenges related to data ownership and control \cite{kayes2017privacy}. Typically, these platforms are governed by a central entity that owns and oversees user data, raising concerns about the potential exploitation of personal information for targeted advertising and other purposes without explicit user consent. Notable examples include the issues associated with Facebook \cite{hull2011contextual,jones2005facebook}. The vulnerability of centralized platforms is evident in the severe risks of identity theft and security breaches resulting from data breaches\cite{gundecha2014user,pierson2012online}.

In contrast, decentralized social networks bring challenges while desiring to empower users with greater control over their data\cite{de2018survey,bahri2018decentralized}. Users handle more direct responsibility for data security and management, necessitating navigation through complexities such as privacy settings and encryption keys. Blockchain technology emerges as a potential solution to addressing these concerns \cite{guidi2020blockchain}. However, in Decentralized Online Social Networks, blockchain implementations often function primarily as a storage mechanism, leading to publicly accessible content, including users' social data. We identified two critical issues in previous works  \cite{datta2010, de2017, choi2020, rahman2019}. Firstly, storing users' content in plaintext requires Trusted nodes to replicate all stored data and authenticate access control lists for each user request. The vulnerability lies in the dependence on Trusted nodes, as compromising even one node could jeopardize the entire system. For instance, a malicious node could be manipulated, stolen, or accept unauthorized user requests. Secondly, users must trust the Trusted nodes, and they cannot verify the accuracy of the content they receive without tampering. Additionally, the replication of storage incurs substantial costs when all Trusted nodes store all users' content.

To address the challenges mentioned above in recent Centralized and Decentralized Online Social Networks, we propose a novel architecture for Decentralized Online Social Networks (DOSN), as illustrated in Figure 5. This architecture leverages a Decentralized Storage Network (DSN), exemplified by Filecoin, to distribute data across multiple nodes, eliminating the need for a central authority. IPFS, combined with Merkle Directed Acyclic Graphs (DAGs), is employed for efficient data representation, facilitating decentralized and redundant data distribution. In the proposed DOSN, user content is distributed across multiple miners, with only the root of the Merkle DAG stored on the blockchain. The Shamir Secret Sharing Scheme is utilized for efficient key management, with miners holding decryption key shares. Access Control Contracts (ACCs) are employed to regulate and manage access to resources, allowing owners to create, update, revoke, and delete access policies. These ACCs ensure secure and authorized interactions within the decentralized network, providing a comprehensive solution to privacy and security concerns in DOSNs. 

The proposed architecture for Decentralized Online Social Networks (DOSNs) represents a substantial and innovative contribution to the field. By leveraging a Decentralized Storage Network (DSN) like Filecoin and integrating IPFS with Merkle Directed Acyclic Graphs (DAGs), the research addresses the limitations of existing decentralized platforms. The distribution of user content across multiple miners, with only the root of the Merkle DAG stored on the blockchain, eliminates the need for a central authority, enhancing data security and privacy. The incorporation of the Shamir Secret Sharing Scheme for key management and Access Control Contracts (ACCs) for resource access regulation further reinforces user control and trust. This proposed solution not only mitigates the vulnerabilities associated with Trusted nodes but also offers a cost-effective and efficient means of data distribution. Overall, the contribution introduces a comprehensive framework that not only tackles the identified challenges but also sets a promising foundation for the development of more secure, user-centric, and decentralized social networks.

To improve our work, it is important to recognize and address its limitations. The present state of research in this field reveals significant gaps and challenges that necessitate attention in future work. Notably, there is a lack of experiments and evaluations on different architectures, hindering a comprehensive understanding of their performance. This lack of empirical data impedes progress and hampers the ability to make informed decisions regarding the optimal selection of architectures. Moreover, the shortage of relevant papers for review poses a challenge in synthesizing a comprehensive body of knowledge. This gap highlights the need for increased contributions to establish a strong foundation for further exploration and analysis. In terms of emerging concepts, their interrelations require careful examination in future research. Exploring and understanding these innovative ideas will be crucial for advancing the field and uncovering potential pathways for innovation. The complexity of the Shamir Secret Sharing Scheme presents a notable challenge, requiring in-depth examination to enhance comprehension and identify potential areas of optimization. Additionally, scalability challenges in decentralized storage systems and the efficiency of smart contract execution justify dedicated research initiatives to address current constraints. A careful examination highlights the lack of primary data regarding the definition of Decentralized Online Social Networks (DOSN), emphasizing the need for empirical research to establish a solid foundation for the conceptualization and understanding of this evolving domain. It is crucial to address these gaps to facilitate progress and ensure the resilience of future developments in the field.

To address the above-mentioned limitations we will implement some future steps. Our research needs a comprehensive exploration of our architecture through experiments and evaluations, aiming to find its complexities and assess its performance across various scenarios.  To enhance our comprehension and strengthen the work, we aim to incorporate new aspects from novel bibliographical approaches. This involves a careful examination of current literature, ensuring that our research stays at the forefront of the most recent theoretical developments.Anticipating the challenges inherent in decentralized storage infrastructure, we will focus on addressing potential scalability issues. This approach involves strategies to enhance the system's capacity and resilience, ensuring its seamless operation even as demands change.In the domain of blockchain technology, we will direct our focus towards optimizing the efficiency of smart contracts. Recognizing their pivotal role in decentralized systems, our research aims to simplify their execution, fostering a more efficient and responsive blockchain environment. Our approach will involve continuous security audits and upgrades for the DOSN architecture. This approach ensures the timely identification and mitigation of emerging vulnerabilities, thereby fortifying the system against potential threats.

\section{Conclusions}
\label{sec:conclusion}

We emphasize the importance of addressing privacy concerns in social networks. By conducting a comprehensive review and proposing an innovative decentralized architecture leveraging blockchain, the study contributes significantly to safeguarding user data. We proposed the DOSN architecture integrates Blockchain technology, Decentralized Storage Networks, and Access Control Smart Contracts, ensuring user control and aligning with GDPR principles. In summary, we present a practical and secure solution that aligns with regulations and establishes a user-centric foundation for the future of digital communication.
%
%
%
%
\bibliographystyle{splncs04}
\bibliography{PrivacyDSN.bib}

\end{document}